\documentclass[english,twocolumn,showpacs]{revtex4-1}
\usepackage[T1]{fontenc}
\usepackage[latin9]{inputenc}
\usepackage{amsmath}
\usepackage{amssymb}
\usepackage{graphicx}
\usepackage{esint}

\makeatletter

\@ifundefined{textcolor}{}
{%
\definecolor{BLACK}{gray}{0}
\definecolor{WHITE}{gray}{1}
\definecolor{RED}{rgb}{1,0,0}
\definecolor{GREEN}{rgb}{0,1,0}
\definecolor{BLUE}{rgb}{0,0,1}
\definecolor{CYAN}{cmyk}{1,0,0,0}
\definecolor{MAGENTA}{cmyk}{0,1,0,0}
\definecolor{YELLOW}{cmyk}{0,0,1,0}
}

\usepackage{hyperref}
\usepackage{xcolor}
\hypersetup{
    colorlinks,
    citecolor=blue,
    filecolor=blue,
    linkcolor=blue,
    urlcolor=blue,
    pdfstartview=FitH
}

\makeatother

\usepackage{babel}
\begin{document}

\title{Generalized Eilenberger theory for Majorana zero-mode-carrying disordered
$p$-wave superconductors}

\author{Hoi-Yin Hui}

\email{hyhui@umd.edu}

\affiliation{Department of Physics, Condensed Matter Theory Center and Joint Quantum
Institute, University of Maryland, College Park, Maryland 20742-4111,
USA}

\author{Jay D. Sau}

\affiliation{Department of Physics, Condensed Matter Theory Center and Joint Quantum
Institute, University of Maryland, College Park, Maryland 20742-4111,
USA}

\author{S. Das Sarma}

\affiliation{Department of Physics, Condensed Matter Theory Center and Joint Quantum
Institute, University of Maryland, College Park, Maryland 20742-4111,
USA}

\date{\today}
\begin{abstract}
Disorder is known to suppress the gap of a topological superconducting
state that would support non-Abelian Majorana zero modes. In this
paper, we study using the self-consistent Born approximation the robustness
of the Majorana modes to disorder within a suitably extended Eilenberger
theory, in which the spatial dependence of the localized Majorana
wave functions is included. We find that the Majorana mode becomes
delocalized with increasing disorder strength as the topological superconducting
gap is suppressed. However, surprisingly, the zero bias peak seems
to survive even for disorder strength exceeding the critical value
necessary for closing the superconducting gap within the Born approximation.
\end{abstract}


\maketitle

\section{Introduction}

Recently, considerable theoretical effort has been put into the search
for localized non-Abelian Majorana modes (MMs) in solid state systems
\cite{Alicea2010Majorana,Leijnse2012Introduction,Beenakker2013Search,Stanescu2013Majorana}.
All proposed systems for the realization of MMs rely on the presence
of $p$-wave superconductivity, either intrinsically \cite{Kitaev2001Unpaired}
or artificially through a careful engineering of heterostructure design
\cite{Fu2008Superconducting,Fu2009Josephson,Sau2010Generic,Sau2010Non-Abelian,Lutchyn2010Majorana,Oreg2010Helical}.
The theoretical progress in this field has sparked substantial experimental
effort in realizing the proposed systems \cite{Mourik2012Signatures,Deng2012Anomalous,Das2012Zero-bias,Churchill2013Superconductor-nanowire,Finck2013Anomalous},
with several measurements reporting the observation of the theoretically
predicted \cite{Sau2010Generic,Sau2010Non-Abelian,Lutchyn2010Majorana}
zero bias peaks (ZBPs) in conductance measurements in semiconductor
nanowires consistent with the existence of zero energy MMs.

However, a compelling and unambiguous signature for the MMs is still
lacking. Among various complications, an unfavorable factor in experiments
is the disorder invariably present in all real experimental samples.
With the possible exception of topological insulator-based heterostructures
\cite{Fu2008Superconducting,Fu2009Josephson}, in which the effects
of disorder are minimal \cite{Hui2014Disorder-induced}, other semiconductor-based
heterostructures \cite{Sau2010Generic,Lutchyn2010Majorana,Oreg2010Helical}
and even the idealized $p$-wave superconductor \cite{Kitaev2001Unpaired}
are all susceptible to disorder, since the Anderson theorem asserting
the insensitivity of superconductivity to ordinary spin-independent
momentum scattering does not in general apply to $p$-wave superconducting
ordering.

The effects of disorder in such topological systems have been previously
investigated \cite{Motrunich2001Griffiths,Akhmerov2011Quantized,Bagrets2012Class,Brouwer2011Probability,Brouwer2011Topological,Liu2012Zero-Bias,Lobos2012Interplay,Pientka2012Enhanced,Sau2012Experimental,DeGottardi2013Majorana,Sau2013Density,Neven2013Quasiclassical,Rainis2013Towards,Adagideli2014Effects}
from a number of different perspectives. One approach to the problem
consists of introducing many realizations of disorder and ensemble-averaging
at the end to extract universal properties \cite{Liu2012Zero-Bias,Pientka2012Enhanced,DeGottardi2013Majorana,Rainis2013Towards,Sau2013Density,Adagideli2014Effects}.
While this approach is more akin to the experimental situation (where
there is only a single realization of disorder at each setup), the
end result of the posterior disorder averaging is mostly numerical
and few analytical statements can be made. On the other hand, previous
attempts of anterior disorder averaging were mostly concerned with
the properties in the bulk \cite{Motrunich2001Griffiths,Akhmerov2011Quantized,Brouwer2011Probability,Brouwer2011Topological,Bagrets2012Class,Lobos2012Interplay,Sau2012Experimental}.
The effects of ensemble-averaged disorder on the end MMs were not
fully investigated.

In this paper, we undertake the task of analyzing the effects of ensemble-averaged
disorder on a topological one-dimensional (1D) system, the idealized
spinless $p$-wave superconducting wire. In particular, we treat the
disorder in the self-consistent Born approximation (SCBA) and investigate
its effects on the spectral properties of the whole system, with an
emphasis on its boundary where the MMs reside. We are generalizing
in the current work the earlier work \cite{Sau2012Experimental} done
by two of the authors in order to analytically assess the disorder
effects on the MMs themselves, not just on the topological superconducting
gap. Of course, this problem has been much studied in the recent MM
literature, but most of this disorder work is carried out purely numerically--a
typical example being our own work presented in Ref.~\onlinecite{Sau2013Density}.
The goal in the current work is to develop an analytical approach
to the problem just as we did for the superconducting gap in Ref.~\onlinecite{Sau2012Experimental}.

To generalize SCBA to inhomogeneous structures, it is convenient to
adopt a formalism similar to the Eilenberger equations \cite{Eilenberger1968Transformation},
but including pair-breaking effects of disorder in the theory. Our
formalism differs from the conventional quasiclassical treatment of
disordered superconductors in two ways. First, unlike the conventional
approach, we consider only weak disorder and do not take the diffusive
limit to derive the Usadel equations \cite{Usadel1970Generalized},
as this would wipe out the spectral gap in the topological system.
It is important to emphasize that the standard Usadel formalism for
ordinary metallic superconductivity, which is the generalization of
the Eilenberger theory to include disorder in the diffusive limit,
is inapplicable to the topological superconducting situation of interest
here sine the system becomes gapless in the Usadel limit. Our current
work provides the appropriate disorder generalization of the quasiclassical
Eilenberger theory for the disordered $p$-wave topological (i.e.
MM-carrying) superconducting system. The standard Usadel theory cannot
be used for topological superconductors since the implicit assumption
of strong disorder scattering in the Usadel theory makes it inapplicable
to the effective $p$-wave superconducting pairing underlying the
MM-carrying topological systems. The second difference of our work
from the conventional treatment is that we do not start by integrating
out the fast-oscillating parts of the Green function, but instead
consider the Green function in a chiral basis and keep all its spatial
dependence. This is possible only for 1D problems, and is essential
to extract the exact spatial dependence of the MM.

We emphasize that the focus of this paper is on the semiconductor
heterostructures in the presence of spin-orbit coupling and spin splitting
\cite{Fu2008Superconducting,Fu2009Josephson,Sau2010Generic,Sau2010Non-Abelian,Lutchyn2010Majorana,Oreg2010Helical}
proposed to realize MMs, all of which have $s$-wave pairing terms
induced by proximity effect through Cooper pairs tunneling from proximate
superconductors. By projecting the Hamiltonians of these systems to
their low-energy subspaces \cite{Lutchyn2010Majorana,Potter2011Engineering,Alicea2012New},
one universally obtains effective $p$-wave superconductors but with
model-specific pairings and scattering strengths {[}see Eq.~(\ref{eq:H})
below{]}. The crucial difference of this semiconductor Majorana nanowire
from an intrinsic $p$-wave superconducting wire is that, since now
the pairing term is proximity-induced, it is both unnecessary and
inappropriate to perform self-consistent theoretical calculations
because there is no intrinsic pairing interaction in the wire itself--the
pairing is induced entirely from the outside through the proximity
effect \cite{Sau2010Robustness}. This issue has already been discussed
in previous numerical \cite{Stanescu2011Majorana} or quasiclassical
\cite{Stanev2014Quasiclassical} investigations of these systems,
and we shall thus take the pairing strength as a fixed parameter without
solving the self-consistent gap equation following all earlier theoretical
works in the literature on this problem.

The paper is organized as follows. In Sec.~\ref{sec:Formalism} we
briefly discuss the formalism we adopt to analyze our current $p$-wave
superconducting system. Then we compare this approach with the result
obtained from SCBA in the bulk in Sec.~\ref{sec:Infinite-Wire}.
Next the effect of disorder on the MMs at the boundary is investigated
in Sec.~\ref{sec:Semi-Infinite-Wire}, where the Eilenberger equations
are solved for a semi-infinite 1D system. In Sec.~\ref{sec:Leakage}
we discuss the manifestation of hybridization between the MM and the
continuum modes in this formalism. Finally, in Sec.~\ref{sec:Conclusion}
we summarize our results.

\section{Formalism\label{sec:Formalism}}

We consider a semi-infinite wire at $x>0$ described by the linearized
Hamiltonian
\begin{eqnarray}
H & = & \sum_{C=R/L}\int_{0}^{\infty}dx\left[-iv_{F}s_{C}\psi_{C}^{\dagger}\partial_{x}\psi_{C}+\Delta s_{C}\psi_{C}\psi_{\bar{C}}\right.\nonumber \\
 &  & \left.+V_{f}\psi_{C}^{\dagger}\psi_{C}+V_{b}\psi_{C}^{\dagger}\psi_{\bar{C}}\right].\label{eq:H}
\end{eqnarray}
Here $v_{F}$ is the Fermi velocity, $\Delta$ is the $p$-wave superconducting
order parameter, and $s_{C}=\pm1$ for $C=R/L$, where $R/L$ denotes
right/left moving electrons. $V_{f/b}$ is the forward/backward scatterings
due to static quenched disorder, assumed to be short ranged in this
work. Coulomb disorder, which might be present in real semiconductor
nanowire systems of experimental interest, will typically be screened
by the surrounding gates, the normal leads, the superconductor, and
by the electrons in the wire themselves leading presumably to short-ranged
elastic disorder. The linearized form of disorder in Eq.~(\ref{eq:H})
is related to the full disorder potential $U$ by
\begin{eqnarray}
V_{f}\left(x\right) & = & \sum_{q\sim0}U_{q}e^{iqx},\\
V_{b}\left(x\right) & = & \sum_{q\sim0}U_{q-2k_{F}}e^{iqx},
\end{eqnarray}
with $k_{F}$ being the Fermi momentum.

The spectral properties of the system are encoded in the Nambu-Gorkov
Green function $G\left(x,t,x',t'\right)=-i\left\langle \mathcal{T}\Psi\left(x,t\right)\Psi^{\dagger}\left(x',t'\right)\right\rangle $,
where $\Psi=\left(\psi_{R},\psi_{L},\psi_{L}^{\dagger},\psi_{R}^{\dagger}\right)^{T}$.
We are interested, following the spirit of the Eilenberger theory
which is being generalized in the current work, in the  Green function
defined as
\begin{equation}
g\left(x,\omega\right)=v_{F}i\lim_{\epsilon\rightarrow0^{+}}\left[G\left(x,x-\epsilon\right)+G\left(x,x+\epsilon\right)\right]\sigma_{3}\tau_{3}\label{eq:g}
\end{equation}
where $\sigma$ and $\tau$ are Pauli matrices acting on the $R/L$
space and particle-hole space, respectively. To extract the density
of states (DOS) from $g$, we note that since the fermion operator
is linearized in the form of $\psi\left(x\right)\simeq\psi_{R}e^{ik_{F}x}+\psi_{L}e^{-ik_{F}x}$,
the Green function of $\psi\left(x\right)$ is related to the Green
function in the chiral basis via
\begin{eqnarray}
G^{(0)}\left(x,x'\right) & \simeq & G_{RR}e^{ik_{F}\left(x-x'\right)}+G_{RL}e^{ik_{F}\left(x+x'\right)}\nonumber \\
 &  & +G_{LR}e^{-ik_{F}\left(x+x'\right)}+G_{LL}e^{-ik_{F}\left(x-x'\right)}.\label{eq:Gdecomp}
\end{eqnarray}
Therefore, the DOS is given by
\begin{eqnarray}
\nu\left(x,\omega\right) & = & \frac{\nu_{0}}{4}\left[{\rm TrRe}\left(g\sigma_{3}\tau_{3}\right)-{\rm TrRe}\left(g\sigma_{-}\tau_{3}\right)e^{2ik_{F}x}\right.\nonumber \\
 &  & \left.+{\rm TrRe}\left(g\sigma_{+}\tau_{3}\right)e^{-2ik_{F}x}\right],\\
 & = & \frac{\nu_{0}}{4}{\rm TrRe}\left(g\sigma_{3}\tau_{3}+2g\sigma_{+}\tau_{3}\cos2k_{F}x\right),\label{eq:DoSg}
\end{eqnarray}
where $\nu_{0}=\frac{1}{\pi v_{F}}$ is the DOS in the normal state
and $\sigma_{\pm}=\frac{1}{2}\left(\sigma_{1}\pm i\sigma_{2}\right)$.
In Eq.~(\ref{eq:DoSg}), the first(second) term contains the slowly
(fast) -oscillating part of the DOS. Conventional derivation of Eilenberger
equations \cite{Eilenberger1968Transformation,Stanev2014Quasiclassical}
effectively ignores the second term. One key aspect of our generalization
is keeping these oscillatory terms which can be done completely analytically
(at least for the 1D problem of current interest).

The equation of motion of $g$ can be derived from the Dyson's equations
of $G$. As both spatial arguments of $G$ are set to $x$, we must
use the two conjugate Dyson's equations:

\begin{subequations}

\label{eqs:DysonG}
\begin{eqnarray}
\left(\omega-H_{{\rm BdG}}-\Sigma\right)G\left(x,y\right) & = & \delta\left(x-y\right),\\
G\left(y,x\right)\left(\omega-H_{{\rm BdG}}-\Sigma\right) & = & \delta\left(x-y\right),
\end{eqnarray}
\end{subequations}where $H_{{\rm BdG}}=-iv_{F}\sigma_{3}\tau_{3}\partial_{x}+\Delta\sigma_{3}\tau_{1}$
and $\Sigma$ is the self-energy due to ensemble-averaged disorder
$V_{f}$ and $V_{b}$ in Eq.~(\ref{eq:H}). Here the derivative acting
on the right is understood as $G\left(y,x\right)i\overleftarrow{\partial_{x}}=-i\partial_{x}G\left(y,x\right)$.
By collecting the terms $\partial_{x}G\left(x,y\right)$ and $\partial_{x}G\left(y,x\right)$,
we have 
\begin{equation}
v_{F}\partial_{x}g=i\left[\omega\sigma_{3}\tau_{3}-i\Delta\tau_{2}-\sigma_{3}\tau_{3}\Sigma,g\right].\label{eq:Eilenberger}
\end{equation}

The derivation of $\Sigma$ in the SCBA is found to be (see Appendix.~\ref{App:disorder-linear})
\begin{eqnarray}
\Sigma\left(x\right) & = & D_{f}\tau_{3}G\tau_{3}\nonumber \\
 &  & +\frac{D_{b}}{2}\left(\sigma_{1}\tau_{3}G\sigma_{1}\tau_{3}+\sigma_{2}\tau_{3}G\sigma_{2}\tau_{3}\right),
\end{eqnarray}
where the disorder strengths $D_{f}$ and $D_{b}$ are defined by\begin{subequations}\label{eqs:disorder-strengths}
\begin{eqnarray}
\left\langle V_{f}\left(x\right)V_{f}\left(x'\right)\right\rangle  & = & D_{f}\delta\left(x-x'\right),\\
\left\langle V_{b}\left(x\right)V_{b}\left(x'\right)\right\rangle  & = & 0,\\
\left\langle V_{b}\left(x\right)V_{b}^{*}\left(x'\right)\right\rangle  & = & D_{b}\delta\left(x-x'\right).
\end{eqnarray}
\end{subequations}

We now resolve Eq.~(\ref{eq:Eilenberger}) into components with the
observation that in the bulk of a clean system $g$ is exactly known
to be
\begin{equation}
g_{{\rm bulk}}=\frac{-i\omega}{\sqrt{\Delta^{2}-\omega^{2}}}\sigma_{3}\tau_{3}-\frac{\Delta}{\sqrt{\Delta^{2}-\omega^{2}}}\sigma_{0}\tau_{2}.\label{eq:gcleanbulk}
\end{equation}
Consider now a situation where $D_{f/b}$ are adiabatically tuned
away from zero in the bulk of the wire. By substituting Eq.~(\ref{eq:gcleanbulk})
in Eq.~(\ref{eq:Eilenberger}), it can be shown that $g$ can only
have six non-zero components:
\begin{eqnarray}
g & = & g_{31}\sigma_{3}\tau_{1}+g_{02}\sigma_{0}\tau_{2}+g_{33}\sigma_{3}\tau_{3}\nonumber \\
 &  & +g_{10}\sigma_{1}\tau_{0}+g_{21}\sigma_{2}\tau_{1}+g_{23}\sigma_{2}\tau_{3},\label{eq:gcomp}
\end{eqnarray}
and their equations of motions are

\begin{subequations}\begin{widetext}

\label{eqs:ODE}
\begin{eqnarray}
v_{F}\partial_{x}g_{31} & = & 2\omega g_{02}+2i\Delta g_{33}+\frac{2i}{\tau}g_{02}g_{33},\label{eq:g1eqm}\\
v_{F}\partial_{x}g_{02} & = & -2\omega g_{31},\label{eq:g2eqm}\\
v_{F}\partial_{x}g_{33} & = & -2i\Delta g_{31}-\frac{2i}{\tau}g_{31}g_{02},\label{eq:g3eqm}\\
v_{F}\partial_{x}g_{10} & = & 2\omega g_{23}-\left(\frac{i}{\tau}-\frac{4i}{\tilde{\tau}}\right)\left(g_{21}g_{31}+g_{23}g_{33}\right),\\
v_{F}\partial_{x}g_{21} & = & 2i\Delta g_{23}+\left(\frac{3i}{\tau}-\frac{4i}{\tilde{\tau}}\right)g_{02}g_{23}+\left(\frac{i}{\tau}-\frac{4i}{\tilde{\tau}}\right)g_{10}g_{31},\\
v_{F}\partial_{x}g_{23} & = & -2\omega g_{10}-2i\Delta g_{21}-\left(\frac{3i}{\tau}-\frac{4i}{\tilde{\tau}}\right)g_{02}g_{21}+\left(\frac{i}{\tau}-\frac{4i}{\tilde{\tau}}\right)g_{10}g_{33},
\end{eqnarray}
\end{widetext}\end{subequations}where we have defined $\tau^{-1}=\pi\nu_{0}D_{b}$
and $\tilde{\tau}^{-1}=\pi\nu_{0}\frac{1}{2}\left(D_{f}+D_{b}\right)$.
Substituting Eq,~(\ref{eq:gcomp}) in Eq.~(\ref{eq:DoSg}), we have
for the DOS
\begin{equation}
\nu\left(x,\omega\right)=\nu_{0}\left({\rm Re}g_{33}-{\rm Im}g_{23}\cos2k_{F}x\right).\label{eq:DOSgcomp}
\end{equation}

To completely formulate the problem, Eqs.~(\ref{eqs:ODE}) must be
supplemented with boundary conditions. In the bulk of the wire $\left(x\rightarrow\infty\right)$,
since the BdG Hamiltonian is diagonal in the $\sigma$ space, the
resultant Green function must also be diagonal in the $\sigma$-space.
This implies that $g_{10}=g_{21}=g_{23}=0$ at $x\rightarrow\infty$.
By setting the spatial derivatives of Eq.~(\ref{eqs:ODE}) to zero,
we also obtain

\begin{subequations}

\label{eqs:BCbulk}
\begin{eqnarray}
\omega g_{02}+i\Delta g_{33}+\frac{i}{\tau}g_{02}g_{33} & = & 0,\\
g_{31} & = & 0,\\
-i\Delta g_{31}-\frac{i}{\tau}g_{31}g_{02} & = & 0
\end{eqnarray}
\end{subequations}at $x\rightarrow\infty$.

To derive the boundary conditions at the end of the wire ($x=0$),
we note that since the fermion operator is linearized as $\psi\left(x\right)=\psi_{R}\left(x\right)e^{ik_{F}x}+\psi_{L}\left(x\right)e^{-ik_{F}x}$,
at the end of wire we have $0=\psi\left(0\right)=\psi_{R}\left(0\right)+\psi_{L}\left(0\right)$.
This translates to the requirement that 
\begin{equation}
\left(\begin{array}{cccc}
1 & 1 & 0 & 0\\
0 & 0 & 1 & 1
\end{array}\right)G\left(0,\epsilon\right)=\left(\begin{array}{cccc}
0 & 0 & 0 & 0\\
0 & 0 & 0 & 0
\end{array}\right).
\end{equation}
Since it follows from the definition of $g$ {[}Eq.~(\ref{eq:g}){]}
and the Dyson's equation for $G$ {[}Eq.~(\ref{eqs:DysonG}){]} that
$\lim_{\epsilon\rightarrow0^{+}}G\left(0,\epsilon\right)=\frac{1}{2iv_{F}}g\left(0\right)\sigma_{3}\tau_{3}+\frac{i}{2v_{F}}\sigma_{3}\tau_{3}$,
we have

\begin{subequations}

\label{eqs:BCend}
\begin{eqnarray}
g_{02} & = & 0,\label{eq:BCend-g02}\\
g_{10} & = & 1,\label{eq:BCend-g10}\\
g_{21} & = & ig_{31},\\
g_{23} & = & ig_{33}
\end{eqnarray}
\end{subequations}at $x=0$. The last condition is also consistent
with the requirement that $\nu\left(0,\omega\right)=0$ {[}c.f. Eq.~(\ref{eq:DOSgcomp}){]}.

Finally we add that since $g^{2}=1$ in the bulk of a clean system
{[}c.f. Eq.~(\ref{eq:gcleanbulk}){]} and from Eq.~(\ref{eq:Eilenberger})
we have $\partial_{x}g^{2}=0$, the normalization $g^{2}=1$ is valid
throughout the whole system. Written in its components,
\begin{equation}
g_{31}^{2}+g_{02}^{2}+g_{33}^{2}+g_{10}^{2}+g_{21}^{2}+g_{23}^{2}=1.\label{eq:normalization}
\end{equation}

We make two remarks before closing the discussion on the formalism.
First, note that Eqs.~(\ref{eqs:ODE}a)-(\ref{eqs:ODE}c) do not
contain the variables $g_{10}$, $g_{21}$, and $g_{23}$. Together
with the boundary conditions Eqs.~(\ref{eqs:BCbulk}) and Eq.~(\ref{eq:BCend-g02}),
$g_{31}$, $g_{02}$, and $g_{33}$ can thus be solved without reference
to the other three variables. These equations have been previously
derived \cite{Eilenberger1968Transformation,Zaitsev1984Quasiclassical}
by first integrating out the fast-oscillating degrees of freedom in
the problem, or equivalently {[}see Eq.~(\ref{eq:Gdecomp}){]} by
assuming that $G$ is always diagonal in $\sigma$-space. We have
seen from Eqs.~(\ref{eqs:BCend}) that this cannot hold true near
the boundary, where the reflection from the end of the wire induces
correlations between left- and right-moving modes. For our current
work, keeping these oscillatory terms, which are always neglected
in the usual Eilenberger theory, is crucial since our interest is
in figuring out the effect of disorder on the MMs which reside at
the boundaries (i.e., at the wire ends of the 1D system).

It can be seen from Eq.~(\ref{eq:DOSgcomp}) that computation of
DOS using $g_{33}$ alone would miss spatially rapid oscillations
near the end of the wire. Indeed, it has been pointed out in Ref.~\onlinecite{Stanev2014Quasiclassical}
that with the reduced set of variables $\left\{ g_{31},g_{02},g_{33}\right\} $,
an oscillatory factor $\left(\propto\cos2k_{F}x\right)$ of the DOS
near the end of the wire is not captured. It is therefore necessary
to solve the whole set of equations (\ref{eqs:ODE}) if a spatial
resolution of the DOS under the Fermi wavelength is desired. However,
in the following sections in this paper, we shall only focus on $\left\{ g_{31},g_{02},g_{33}\right\} $
for simplicity.

Lastly we adopt this formalism to the case of conventional $s$-wave
superconductivity, with the linearized Hamiltonian
\begin{eqnarray}
H_{0} & = & \sum_{C,\sigma}\int_{0}^{\infty}dx\left[-iv_{F}s_{C}\psi_{C\sigma}^{\dagger}\partial_{x}\psi_{C\sigma}+\Delta_{s}\psi_{C\sigma}\psi_{\bar{C},\bar{\sigma}}\right.\nonumber \\
 &  & \left.+V_{f}\psi_{C\sigma}^{\dagger}\psi_{C\sigma}+V_{b}\psi_{C\sigma}^{\dagger}\psi_{\bar{C},\sigma}\right],\label{eq:HS}
\end{eqnarray}
where only non-magnetic disorder $V_{f/b}$ is considered here. Repeating
the above procedures in solving for $\partial_{x}g^{(s)}$ and then
decomposing $g_{s}^{(s)}$ as
\begin{eqnarray}
g^{(s)} & = & g_{01}^{(s)}\sigma_{0}\tau_{1}+g_{32}^{(s)}\sigma_{3}\tau_{2}+g_{33}^{(s)}\sigma_{3}\tau_{3}\nonumber \\
 &  & +g_{10}^{(s)}\sigma_{1}\tau_{0}+g_{22}^{(s)}\sigma_{2}\tau_{2}+g_{23}^{(s)}\sigma_{2}\tau_{3},
\end{eqnarray}
we reach the following set of differential equations:

\begin{subequations}\begin{widetext}

\label{eqs:ODE-S}
\begin{eqnarray}
v_{F}\partial_{x}g_{01}^{(s)} & = & 2\omega g_{32}^{(s)}+2i\Delta_{s}g_{33}^{(s)},\\
v_{F}\partial_{x}g_{32}^{(s)} & = & -2\omega g_{01}^{(s)}-\frac{2i}{\tau}g_{01}^{(s)}g_{33}^{(s)},\\
v_{F}\partial_{x}g_{33}^{(s)} & = & -2i\Delta_{s}g_{01}^{(s)}+\frac{2i}{\tau}g_{01}^{(s)}g_{32}^{(s)},\\
v_{F}\partial_{x}g_{10}^{(s)} & = & 2\omega g_{23}^{(s)}-2i\Delta_{s}g_{22}^{(s)}-\left(\frac{i}{\tau}-\frac{4i}{\tilde{\tau}}\right)\left(g_{22}^{(s)}g_{32}^{(s)}+g_{23}^{(s)}g_{33}^{(s)}\right),\\
v_{F}\partial_{x}g_{22}^{(s)} & = & 2i\Delta_{s}g_{10}^{(s)}-\left(\frac{3i}{\tau}-\frac{4i}{\tilde{\tau}}\right)g_{01}^{(s)}g_{23}^{(s)}+\left(\frac{i}{\tau}-\frac{4i}{\tilde{\tau}}\right)g_{10}^{(s)}g_{32}^{(s)},\\
v_{F}\partial_{x}g_{23}^{(s)} & = & -2\omega g_{10}^{(s)}+\left(\frac{3i}{\tau}-\frac{4i}{\tilde{\tau}}\right)g_{01}^{(s)}g_{22}^{(s)}+\left(\frac{i}{\tau}-\frac{4i}{\tilde{\tau}}\right)g_{10}^{(s)}g_{33}^{(s)},
\end{eqnarray}
\end{widetext}\end{subequations}and the boundary conditions that

\begin{subequations}

\label{eqs:BCbulk-S}
\begin{eqnarray}
g_{10}^{(s)}=g_{22}^{(s)}=g_{23}^{(s)} & = & 0,\\
\omega g_{32}^{(s)}+i\Delta_{s}g_{33}^{(s)} & = & 0,\\
-\omega g_{01}^{(s)}-\frac{i}{\tau}g_{01}^{(s)}g_{33}^{(s)} & = & 0,\\
-i\Delta_{s}g_{01}^{(s)}+\frac{i}{\tau}g_{01}^{(s)}g_{32}^{(s)} & = & 0
\end{eqnarray}
\end{subequations}at $x\rightarrow\infty$ and\begin{subequations}

\label{eqs:BCend-S} 
\begin{eqnarray}
g_{01}^{(s)} & = & 0,\label{eq:BCend-S-g01}\\
g_{10}^{(s)} & = & 1,\\
g_{22}^{(s)} & = & ig_{32}^{(s)},\\
g_{23}^{(s)} & = & ig_{33}^{(s)}
\end{eqnarray}
\end{subequations}at $x=0.$ A normalization condition similar to
Eq.~(\ref{eq:normalization}) also holds:
\begin{equation}
\left(g_{01}^{(s)}\right)^{2}+\left(g_{32}^{(s)}\right)^{2}+\left(g_{33}^{(s)}\right)^{2}+\left(g_{10}^{(s)}\right)^{2}+\left(g_{22}^{(s)}\right)^{2}+\left(g_{23}^{(s)}\right)^{2}=1.\label{eq:normalization-S}
\end{equation}
We observe that $g_{01}^{(s)}$, $g_{32}^{(s)}$, and $g_{33}^{(s)}$
can be solved from Eqs.~(\ref{eqs:ODE-S}a)-(\ref{eqs:ODE-S}c),
Eqs.~(\ref{eqs:BCbulk-S}b)-(\ref{eqs:BCbulk-S}d) and Eq.~(\ref{eq:BCend-S-g01})
independent of the remaining components.

\section{DOS in the Bulk\label{sec:Infinite-Wire}}

\begin{figure}
\begin{centering}
\includegraphics[width=0.95\columnwidth]{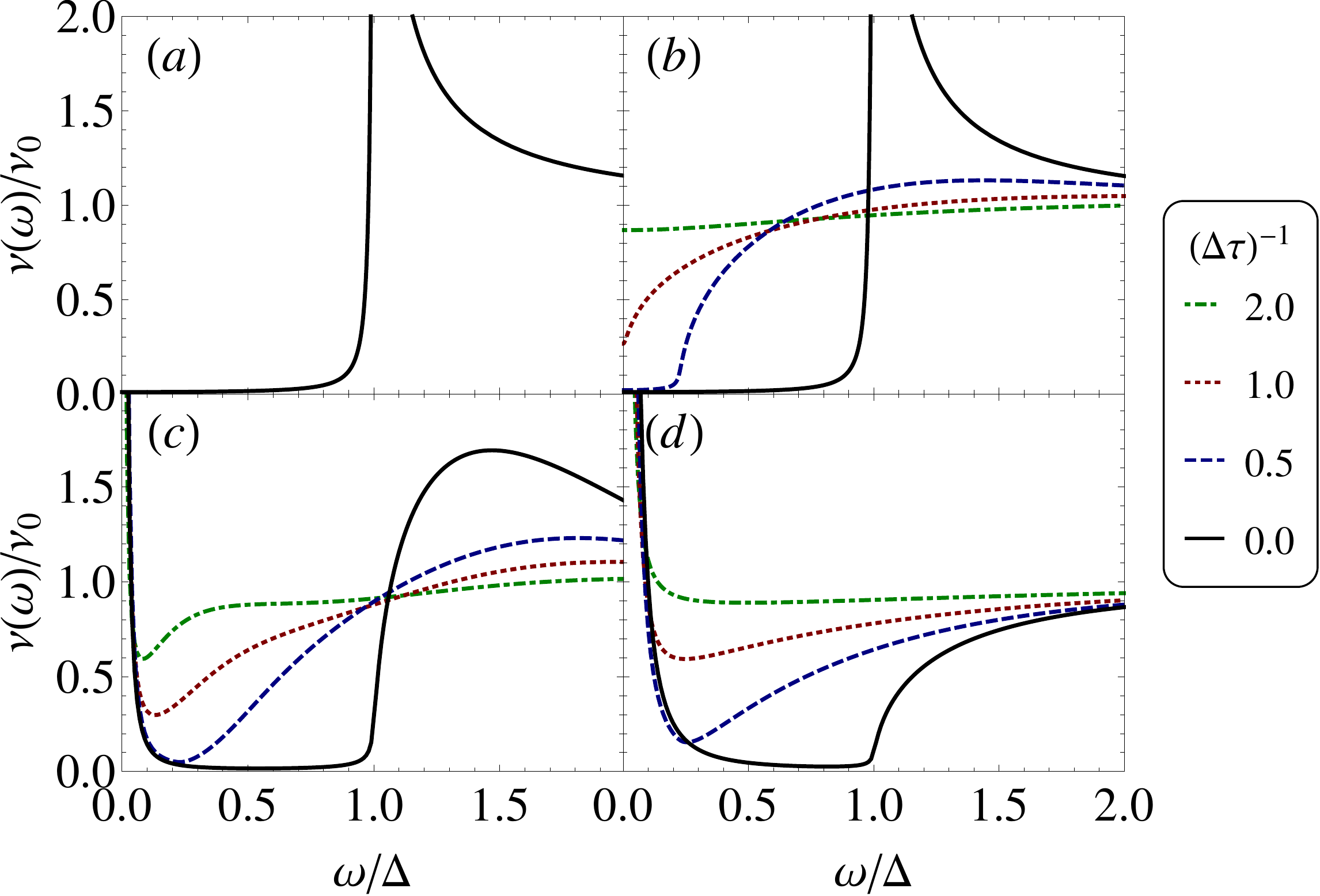}
\par\end{centering}

\caption{(Color online) (a) DOS for a semi-infinite $s$-wave superconducting
wire. Note the result is position-independent, and is not affected
by disorder. (b)-(d) The DOS of a semi-infinite $p$-wave superconducting
wire, from the clean limit $\tau^{-1}=0$ to a heavily disordered
case $\tau^{-1}=2\Delta$ at (b) $x\rightarrow\infty$ (in the bulk),
(c) $x=\xi_{0}$, and (d) $x=0$, respectively. For all plots, the
energy spectra are broadened by $\eta=0.01\Delta$.\label{fig:LDOS-semi-inf}}
\end{figure}

Equations (\ref{eqs:ODE}) can be understood as a generalization of
the SCBA to spatially inhomogeneous structures. Before we utilize
it to investigate into such structures, however, it is instructive
to show that our formalism in the bulk indeed reduces to the SCBA
result obtained earlier by two of the authors \cite{Sau2012Experimental}.

We first consider the simpler case of an $s$-wave superconducting
wire, for which Eqs.~(\ref{eqs:BCbulk-S}) and Eq.~(\ref{eq:normalization-S})
are solved by

\begin{subequations}

\label{eq:SolBulk-S}
\begin{align}
g_{01}^{(s)} & =0,\\
g_{32}^{(s)} & =\frac{-\Delta_{s}}{\sqrt{\Delta_{s}^{2}-\omega^{2}}},\\
g_{33}^{(s)} & =\frac{-i\omega}{\sqrt{\Delta_{s}^{2}-\omega^{2}}},
\end{align}
\end{subequations}independent of the disorder parameter $\tau$.
Therefore, in the $s$-wave case, the DOS in the bulk is
\begin{equation}
\nu_{s}\left(\omega\right)=\nu_{0}{\rm Re}\left[g_{33}^{(s)}\left(\omega\right)\right]=\nu_{0}\frac{\omega}{\sqrt{\omega^{2}-\Delta_{s}^{2}}}\theta\left(\omega-\Delta_{s}\right),\label{eq:SbulkDOS}
\end{equation}
plotted in Fig.~\ref{fig:LDOS-semi-inf}(a), and is unaffected by
disorder as required by Anderson's theorem \cite{Maki1969Superconductivity}.

For the case of $p$-wave superconducting wire in which Anderson's
theorem is not applicable, a suppression of the gap by disorder is
expected. To show this, note that Eqs.~(\ref{eqs:BCbulk}) and Eq.~(\ref{eq:normalization})
are solved by

\begin{subequations}

\label{eq:SolBulk}
\begin{align}
g_{31} & =0,\\
g_{02} & =\frac{-\Delta}{\sqrt{\Delta^{2}-\tilde{\omega}^{2}}},\\
g_{33} & =\frac{-i\tilde{\omega}}{\sqrt{\Delta^{2}-\tilde{\omega}^{2}}},\label{eq:g33bulk}
\end{align}
\end{subequations}where $\tilde{\omega}$ satisfies $\tilde{\omega}=\omega+\frac{i\tilde{\omega}}{\tau\sqrt{\tilde{\omega}^{2}-\Delta^{2}}}$.
This is seen to be identical to the SCBA result of $\tilde{\omega}=\omega+\left(D_{f}+D_{b}\right)\pi\nu_{0}\frac{i\tilde{\omega}}{\sqrt{\tilde{\omega}^{2}-\Delta^{2}}}$,
by noting that for point scatterers $D_{f}=D_{b}$. Figure \ref{fig:LDOS-semi-inf}(b)
is a plot of the DOS evaluated by Eq.~(\ref{eq:DOSgcomp}), for a
number of disorder strengths. The bulk gap is seen to close at about
$\left(\Delta\tau\right)^{-1}=1$. In fact, it can be shown that Eq.~(\ref{eq:SolBulk})
results in a degradation of the spectral gap in the form of \cite{Maki1969Superconductivity}
$E_{{\rm gap}}=\Delta\left[1-\left(\Delta\tau\right)^{-2/3}\right]^{3/2}$,
and eventually destroys the gap for $\tau^{-1}>\Delta$. The influence
of this effect on the MM located at the boundary of the wire is the
focus of the following sections.

\section{DOS Near the End of the Wire\label{sec:Semi-Infinite-Wire}}

We now investigate the effect of ensemble-averaged disorder on the
DOS near the boundary $x=0$. Before considering the case of $p$-wave
superconducting wire in which a MM is present, for the sake of comparison
and illustration, we first review the case of a conventional $s$-wave
superconducting wire in the current formalism. We note that the solution
in the bulk given by Eq.~(\ref{eq:SolBulk-S}) already satisfies
the boundary conditions at the end of the wire {[}Eq.~(\ref{eqs:BCend-S}){]}.
Therefore, the DOS is uniform throughout the whole wire, and Fig.~\ref{fig:LDOS-semi-inf}(a)
is independent of the distance from the boundary. Thus, as expected,
the boundaries of the 1D system or the wire ends do not produce any
nontrivial effects for $s$-wave superconducting wires.

\begin{figure}
\begin{centering}
\includegraphics[width=0.9\columnwidth]{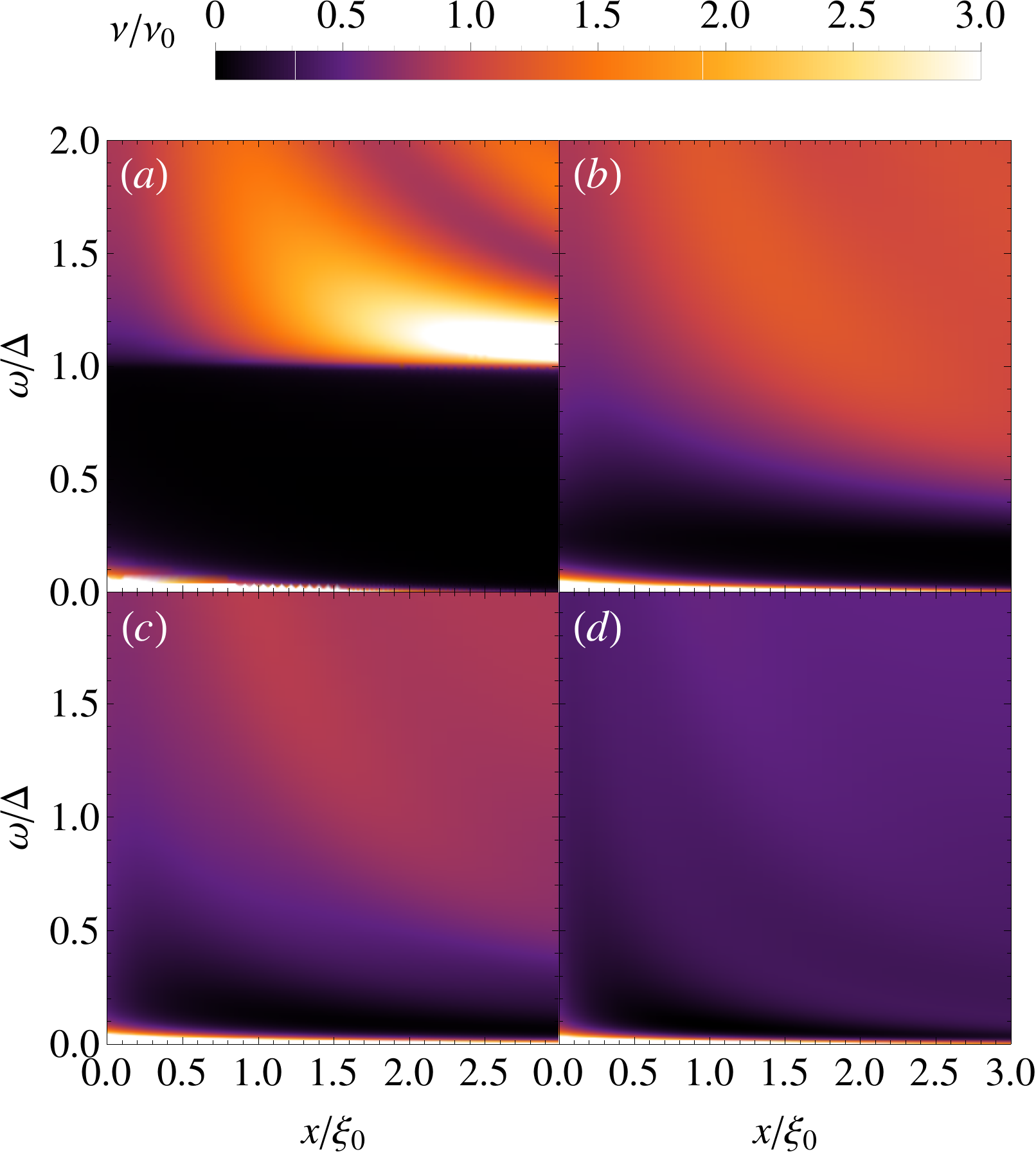}
\par\end{centering}

\caption{(Color online) The DOS $\nu\left(x,\omega,\eta\right)=\nu_{0}{\rm Re}\left[g_{33}\left(x,\omega+i\eta\right)\right]$
plotted as a function of position $x$ (in units of $\xi_{0}=v_{F}/\Delta$)
and energy $\omega$ (in units of $\Delta$), where $x$ is measured
from the end of the wire and $\eta=0.01\Delta$ is the broadening
parameter. The four panels correspond to disorder strengths (a) $\tau^{-1}=0$,
(b) $\tau^{-1}=0.5\Delta$, (c) $\tau^{-1}=\Delta$, and (d) $\tau^{-1}=2\Delta$.
When the system is clean, the salient features are the zero-energy
peak localized at the end and a pristine bulk gap. As disorder is
introduced, the bulk gap shrinks and the singularity is smeared out,
homogenizing the DOS of the whole system, but the zero-energy peak
at the end of the wire is still visible even at strong disorder. \label{fig:contour-semi-inf}}
\end{figure}

In the more nontrivial case of $p$-wave superconductor, the solution
in the bulk Eq.~(\ref{eq:SolBulk}) cannot satisfy the boundary condition
at the end {[}Eq.~(\ref{eq:BCend-g02}){]} and thus Eqs.~(\ref{eqs:ODE})
must be solved directly. Without disorder, the solution is\cite{Stanev2014Quasiclassical}
\begin{eqnarray}
g_{31} & = & \frac{\Delta e^{-2x\sqrt{\Delta^{2}-\omega^{2}}/v_{F}}}{\omega},\\
g_{02} & = & \frac{\Delta\left(e^{-2x\sqrt{\Delta^{2}-\omega^{2}}/v_{F}}-1\right)}{\sqrt{\Delta^{2}-\omega^{2}}},\\
g_{33} & = & i\frac{\Delta^{2}e^{-2x\sqrt{\Delta^{2}-\omega^{2}}/v_{F}}-\omega^{2}}{\omega\sqrt{\Delta^{2}-\omega^{2}}},
\end{eqnarray}
and the other components of $g$ can also be solved analytically but
we shall not state them here as we are ignoring variations in the
length scale of $k_{F}^{-1}$. Note that $g_{31}$ is odd in frequency,
indicating an odd-frequency $s$-wave pairing present near the boundary
\cite{Stanev2014Quasiclassical}. The close relation between the odd-frequency
pairing and MMs has been emphasized in the literature \cite{Tanaka2007Theory,Asano2013Majorana}.

With nonzero disorder, the problem must be solved numerically. Figures
\ref{fig:LDOS-semi-inf}(b)-\ref{fig:LDOS-semi-inf}(d) show the DOS
given by Eq.~(\ref{eq:DOSgcomp}), evaluated in the bulk, at $x=\xi_{0}$
and $x=0$ for a number of disorder strengths. For the same choice
of disorder strengths, the contour plots of the DOS are shown in Fig.~\ref{fig:contour-semi-inf}.
In a clean wire, a singularity in DOS is present at the gap edge $\left(\omega=\Delta\right)$.
This singularity is absent at the end of the wire, where instead a
single zero-energy MM is present. As disorder is introduced, the DOS
throughout the system is homogenized, with the DOS singularity smoothened
and the bulk gap suppressed. As the disorder strength is increased
beyond the bulk-gap closing point of $\tau^{-1}=\Delta$, the continuum
states begin to hybridize with the MM, but the ZBP is distinctly visible
even under strong disorder of $\tau^{-1}=2\Delta$, where in the bulk
the DOS becomes almost flat. It might be of interest to note that
at strong disorder a suppression of the DOS at $\omega\gtrsim0$ is
present only at $x\sim\xi_{0}$, but is absent either in the bulk
or at the end of the wire. This can be understood as the MM is centered
at the end, its hybridization with the continuum states is the strongest
there too.

We point out as an aside that the somewhat surprising continued survival
of the zero mode even beyond the disorder-induced gap closing point
obtained in our current formal semiclassical theory has also been
seen in the direct numerical simulations carried out by two of us
recently \cite{Sau2013Density}. This indicates that the end MMs are
very robust and exist even in the gapless $p$-wave superconducting
phase, which might be consistent with the experimental observations
where the ZBP exists even when there is no obvious gap signature in
the tunneling spectrum.

\section{Change of Majorana Localization Length Under Disorder}

In a clean system the MM is exponentially localized with a decay length
equal to the coherence length $l_{{\rm loc}}=\xi_{0}=v_{F}/\Delta$.
One expects disorder to modify this localization length, which should
diverge as disorder destroys the topological phase\cite{Motrunich2001Griffiths}.
On the one hand, the suppression of the spectral gap seems to suggest
a longer decay length if it is substituted into the formula $l_{{\rm loc}}=v_{F}/E_{{\rm gap}}$.
On the other hand, in the case of $s$-wave superconductors, the coherence
length of a strongly disordered system is shortened to $\xi_{{\rm dis}}\approx v_{F}\sqrt{\tau/\Delta}$,
which suggests a shorter decay length if the formula $l_{{\rm loc}}=\xi_{{\rm dis}}$
is to be trusted. Equations (\ref{eqs:ODE}) allow for a quantitative
investigation of the problem.

\begin{figure}
\begin{centering}
\includegraphics[width=0.85\columnwidth]{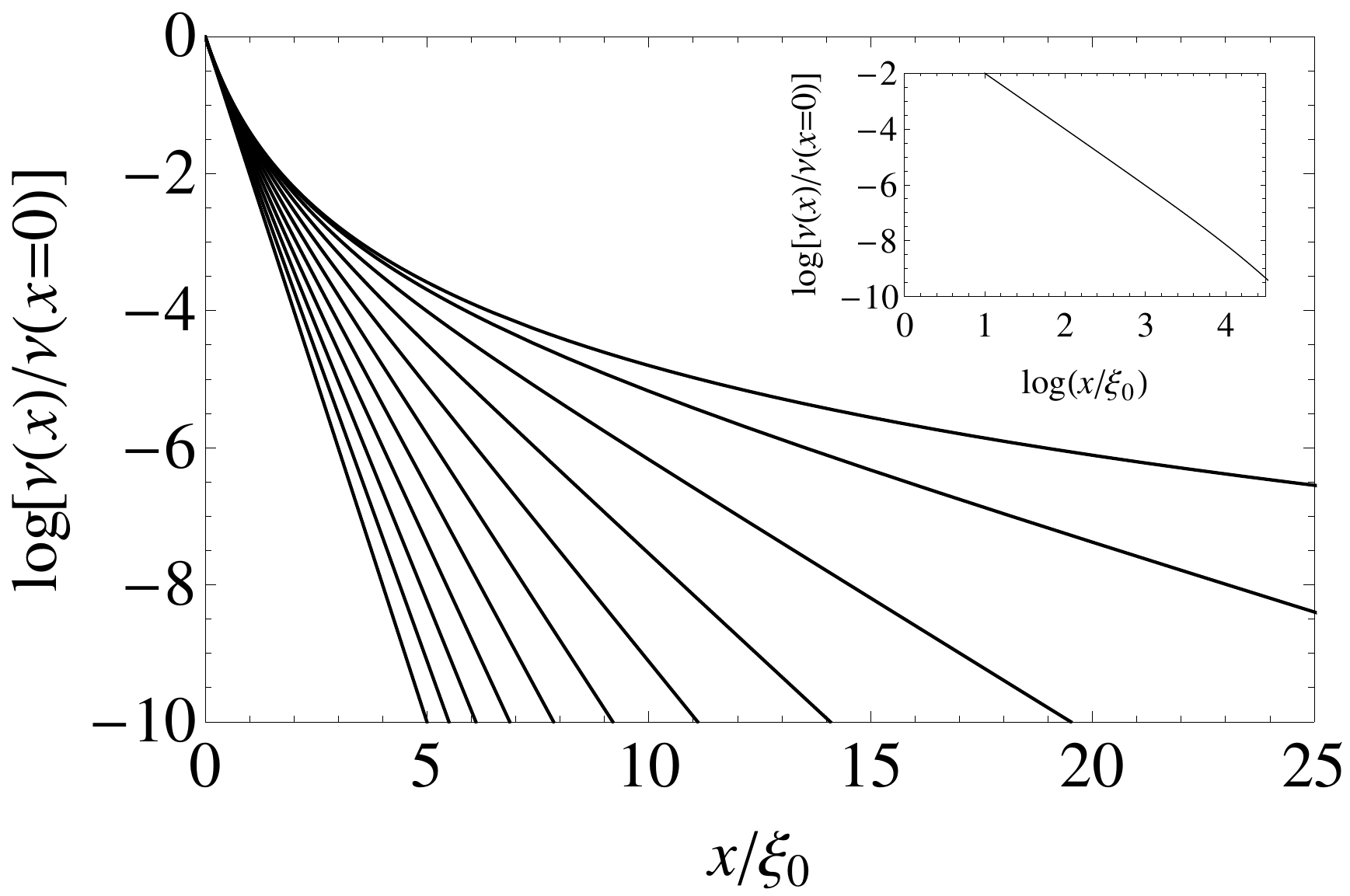}
\par\end{centering}

\caption{Log-linear plot of zero-energy DOS $\nu\left(x,\omega=0\right)$ as
a function of distance $x$ measured from the end of a $p$-wave superconducting
wire, with oscillations of length scale $k_{F}^{-1}$ ignored. The
steepest line corresponds to clean case $\tau^{-1}=0$ where the MM
is most localized. The least steep line corresponds to the critical
disorder strength $\tau^{-1}=\Delta$ where the bulk gap closes. The
intermediate lines are sampled at equally spaced $\tau^{-1}$ with
a step size of $\delta\left(\tau^{-1}\right)=0.1\Delta$. The inset
shows the last curve corresponding to $\tau^{-1}=\Delta$ in log-log
scale. Its slope is approximately $-2$. \label{fig:Majorana-Profile}}
\end{figure}

The decay length is extracted in the following way. The DOS is related
to the Green function by $\nu\left(x,\omega\right)\propto\sum_{n}\frac{\psi_{n}\left(x\right)\psi_{n}^{*}\left(x\right)}{\omega-E_{n}+i\delta}$
where the summation is over all eigenmodes with energies $E_{n}$.
Therefore, a localized zero-energy MM with wave function of the form
$\sim e^{-x/\xi}$ will result in a decay of the DOS as $\nu\left(x,\omega=0\right)\sim e^{-2x/\xi}$,
provided that the bulk gap is finite. Note that it is convenient to
ignore the fast-oscillating DOS contributed by $g_{23}$ in Eq.~(\ref{eq:DOSgcomp}).

In Fig.~\ref{fig:Majorana-Profile} we plot the the zero-energy DOS
$\nu\left(x,\omega=0\right)$ in log scale, for a range of disorder
strength $\tau^{-1}$ up to the critical strength where the bulk gap
closes. For the clean limit $\tau^{-1}=0$, the plot is linear with
a slope of $\frac{-2}{\xi_{0}}$, as expected since the MM is localized
with a decay length of $\xi_{0}$. When disorder is increased, the
slope decreases in magnitude and the curve deviates from a linear
behavior. As the strength is increased to the critical gap-closing
value $\left(\tau^{-1}=\Delta\right)$, the decay ceases to be exponential
and becomes power-law in nature, as is clear from the linearity of
the curve in the log-log plot shown in the inset of Fig.~\ref{fig:Majorana-Profile}.
A linear fit through the log-log plot shows that the decay of the
ZBP is a power law with a behavior of $x^{-1}$.

\begin{figure}
\begin{centering}
\includegraphics[width=0.8\columnwidth]{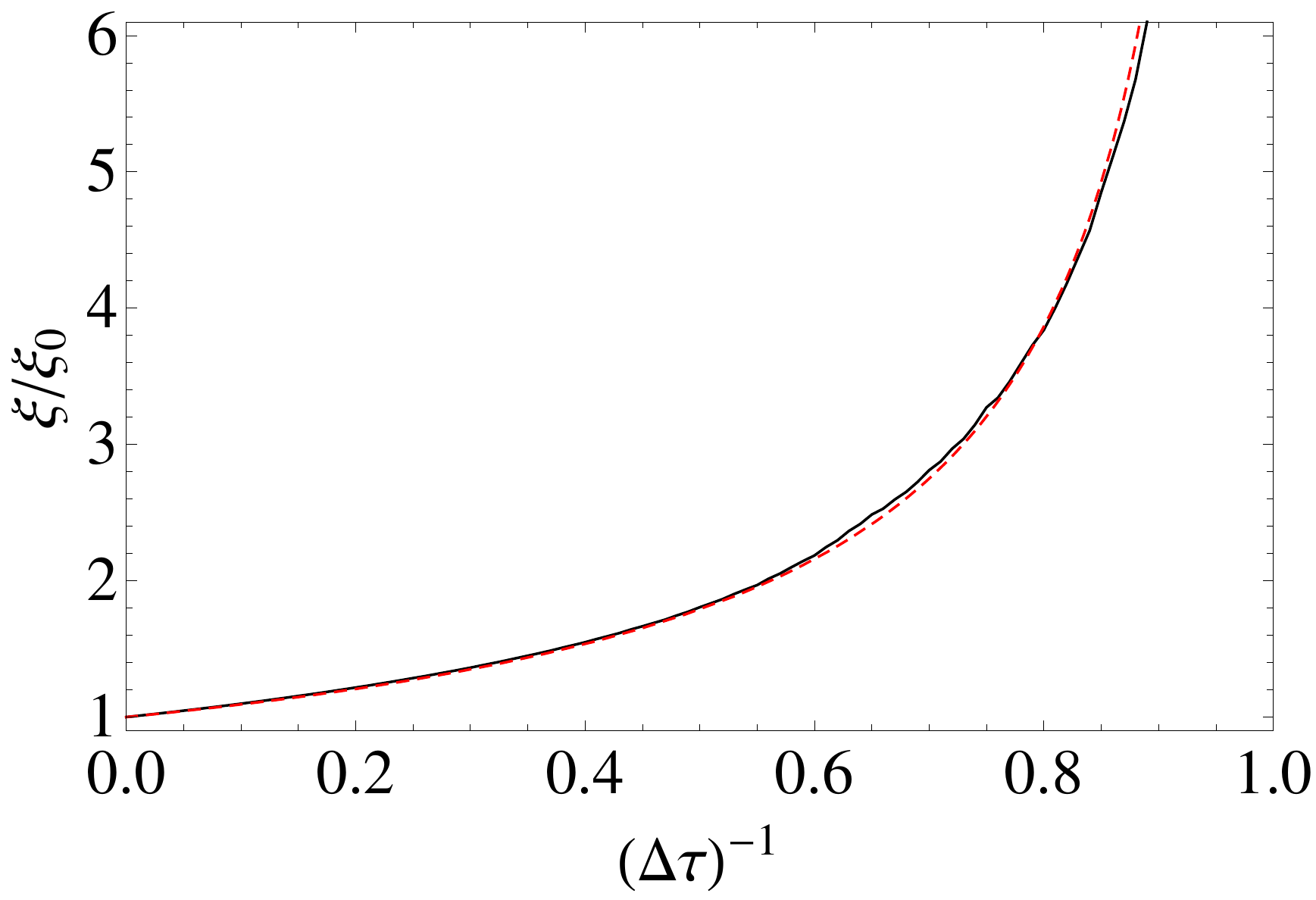}
\par\end{centering}

\caption{(Color online) Plot of the localization length of the MM as a function
of disorder strength. The black solid line shows the numerical values
extracted from Fig.~\ref{fig:Majorana-Profile} by fitting the tails
of the curves (at $\log\frac{\nu\left(x\right)}{\nu\left(x=0\right)}<-6$)
with straight lines. Note that the result is meaningful only for weak
disorder $\left(\tau^{-1}\lesssim\Delta\right)$ where the corresponding
the curve in Fig.~\ref{fig:Majorana-Profile} is approximately linear.
The red dashed line is the best-fit line of a power-law form of $\frac{\xi}{\xi_{0}}=\left[1-\left(\Delta\tau\right)^{-1}\right]^{-0.84}$.
\label{fig:MMfit}}
\end{figure}

To be more quantitative, the decay length $\xi$ of the Majorana mode
could be crudely estimated from Fig.~\ref{fig:Majorana-Profile},
in the weak disorder limit (roughly when $\tau^{-1}\lesssim\Delta$)
where the curves are approximately linear, by fitting the curves with
straight lines. We compute the slope $m$ of the best-fit line of
the tail of each curve in Fig.~\ref{fig:Majorana-Profile} and extract
the estimated decay length $\xi$ of the Majorana mode by $\xi\sim\frac{-2}{m}$,
with the results shown in Fig.~\ref{fig:MMfit}. For the purpose
of completeness, Fig.~\ref{fig:MMfit} is presented with disorder
ranging from zero to the gap-closure limit $\left(\tau^{-1}=\Delta\right)$,
but it should be cautioned that near the gap-closure limit the notion
of ``decay length'' is meaningless as the decay behavior shows a crossover
from exponential to power-law. To understand the nature of the divergence
at $\tau^{-1}=\Delta$, we fit the curve with a power-law function
and obtain $\frac{\xi}{\xi_{0}}\simeq\left[1-\left(\Delta\tau\right)^{-1}\right]^{-0.84}$.
Figure \ref{fig:MMfit} shows that this empirical form captures the
variations of decay length very well.

\section{Leakage of the Majorana Mode\label{sec:Leakage}}

The zero-energy MM appears to persist even after the gap closes within
our  formalism. More precisely, the DOS at the boundary $\nu\left(x=0,\omega\right)$
has a pole at $\omega=0$ for any finite values of $\Delta$ and $\tau$.
This fact could be derived directly from Eqs.~(\ref{eqs:ODE}a)-(\ref{eqs:ODE}c)
with a perturbative treatment in $\Delta$ (see Appendix~\ref{App:pole}).
As we know from the case of the clean wire that the divergence at
zero-energy comes from a single MM, we fit the DOS near the end of
the wire and near zero energy with a Lorentzian form:

\begin{equation}
\nu_{\tau}\left(x,\omega,\eta\right)\sim\frac{1}{2\pi}Z_{\tau}\left(x\right)\frac{\eta}{\omega^{2}+\eta^{2}}+\nu_{{\rm reg}},\label{eq:nudis}
\end{equation}
where $Z_{\tau}\left(x\right)$ is a the fitting parameter and the
subscript $\tau$ indicates the dependence on disorder strength. $\eta$
is an artificial broadening parameter and $\nu_{{\rm reg}}$ is the
part of the DOS that remains non-divergent as $\eta,\omega\rightarrow0$,
contributed from the other delocalized modes in the system.

On the other hand, we know that if the DOS is contributed by a single
mode $\psi_{0}$, its exact form is
\begin{equation}
\nu^{(0)}\left(x,\omega,\eta\right)=\frac{1}{2\pi}\sum_{\lambda}\left|\psi_{0\lambda}\left(x\right)\right|^{2}\frac{\eta}{\omega^{2}+\eta^{2}},\label{eq:nu0}
\end{equation}
where the summation $\Sigma_{\lambda}$ is over the four-component
BdG spinor. Comparing Eqs.~(\ref{eq:nudis}) and (\ref{eq:nu0}),
it is seen that the spectral weight defined as $Z_{\tau}=\int_{0}^{\infty}Z_{\tau}\left(x\right)dx$
is normalized to unity provided that the MM is not hybridized with
other modes.

\begin{figure}
\begin{centering}
\includegraphics[width=0.8\columnwidth]{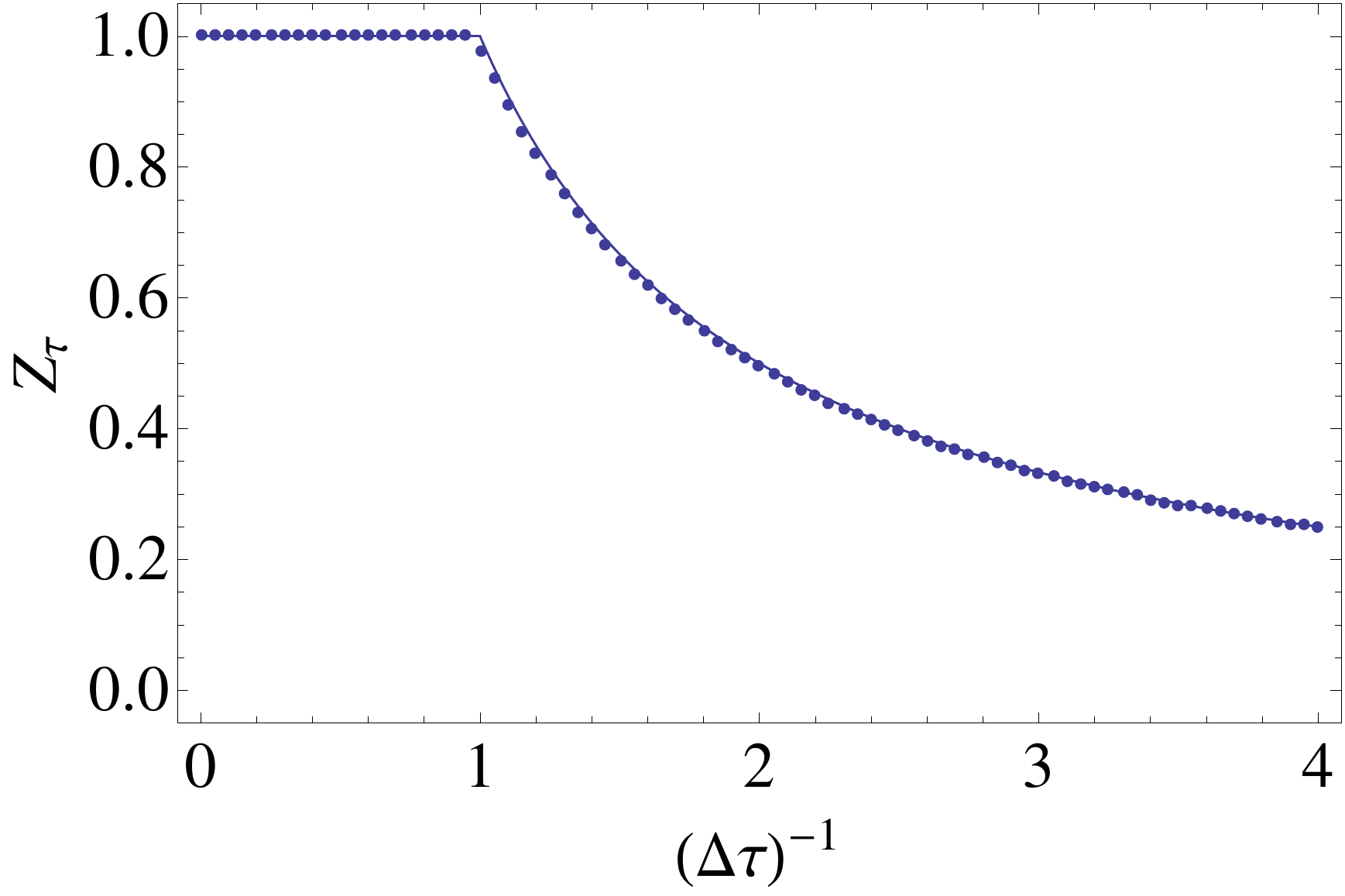}
\par\end{centering}

\caption{(Color online) Plot of the spectral weight $Z_{\tau}$ (defined in
the text) of the zero-energy end mode against disorder strength. The
dots show the values obtained from the numerical solution of Eqs.~(\ref{eqs:ODE})
for a range of disorder strengths. The solid line plots the empirical
formula Eq.~(\ref{eq:empZ}). \label{fig:Z}}
\end{figure}

Figure \ref{fig:Z} shows the variations of $Z_{\tau}$ as the strength
of disorder is changed. For $\Delta\tau\geq1$, $Z_{\tau}$ remains
around unity, which is expected as the bulk gap is not closed and
the zero-energy MM remains exponentially localized and protected by
the spectral gap (and therefore of unit spectral weight). As disorder
is increased beyond the strength where the bulk gap closes, $Z_{\tau}$
starts to decrease below unity. This reduction in the spectral weight
can be understood as a consequence of the hybridization between the
continuum modes in the bulk and the MM. Interestingly, the dependence
of $Z_{\tau}$ on disorder can be captured almost perfectly with the
empirical formula
\begin{equation}
Z_{\tau}=\begin{cases}
1, & \Delta\tau\geq1\\
\Delta\tau, & \Delta\tau<1.
\end{cases}\label{eq:empZ}
\end{equation}

We note that Eq.~(\ref{eq:empZ}) indicates a continuous decrease
of the MM spectral weight from unity in the topologically gapped situation
to a small, but not necessarily vanishingly small, value in the gapless
phase. This robustness of the MM spectral weight even in the presence
of fairly strong disorder (which completely closers the bulk topological
gap) may be the reason for the existence of the ZBP in nanowires which
do not necessarily have very high mobilities or obvious superconducting
gaps.

\section{Conclusion\label{sec:Conclusion}}

In this paper we have derived a theory for a disordered $p$-wave
superconductor in 1D, with the effects of disorder incorporated by
SCBA. Our theory is thus the $p$-wave generalization of the Eilenberger
theory to 1D systems with the explicit inclusion of disorder. A brief
comparison with previous works is in order. Reference \onlinecite{Neven2013Quasiclassical}
applied the Eilenberger equations to a spin-orbit coupled wire with
proximity-induced Zeeman term and superconductivity, but the disorder
was introduced \emph{after} the Eilenberger equations were obtained
and explicit disorder-averaging was performed numerically. Reference
\onlinecite{Stanev2014Quasiclassical} adopted the Eilenberger equations
to the same system investigated by us, but the emphasis was put on
the analysis of proximity effect and \emph{no} disorder was introduced.
Moreover, short-length-scale fluctuations in the DOS were explicitly
ignored in Reference \onlinecite{Stanev2014Quasiclassical}. Our study
differs from these works in that disorder is incorporated by SCBA
in the Eilenberger equations, and spatial fluctuations of the DOS
of the order of Fermi wavelength is retained. In fact, the inclusion
of both disorder and spatial fluctuations are the main features of
our theory distinguishing it from earlier works in the literature.

We applied our formalism to a semi-infinite $p$-wave superconducting
wire, and found that the gap of the system in the bulk is suppressed
by disorder in a way consistent with previous studies. We then focused
on the MM located at the end of the wire. We found that with the bulk
gap being suppressed, the localization length of the MM increases,
and diverges when the gap vanishes. In this process, the localization
behavior of the MM changes from exponential to a power-law decay.
We also pointed out an unusual feature of the MM under disorder in
this formalism: the DOS shows a divergence at zero-energy at the end
of wire even at strong disorder. This is contradictory to the fact
that the MM should hybridize with the continuum modes and its spectrum
should broaden. However, we can still extract certain manifestations
of this hybridization within this formalism--the spectral weight of
the MM decreases after the bulk gap is closed, showing a ``leakage''
of the MM to the continuum. It is interesting that we find that some
vestiges (``Majorana ghosts'') of the MMs survive strong disorder
and continue showing up in the zero-energy DOS even when the $p$-wave
system has become essentially a gapless system due to disorder.

The results from SCBA appear qualitatively consistent with numerical
solutions of the DOS \cite{Sau2013Density} near the end. In these
studies the ZBP, which starts as a sharp Majorana peak, decreases
in height and broadens out into a peak resulting from Griffiths singularities
\cite{Motrunich2001Griffiths} that is consistent with the class-D
symmetry of the system \cite{Neven2013Quasiclassical}. In contrast
to the more exact results where the ZBP is found to broaden into a
power-law singularity, we find that the ZBP stays sharp near zero
energy while reducing in spectral weight. This discrepancy is not
unexpected since the SCBA is a mean-field theory and cannot possibly
describe critical fluctuations. Furthermore, we cannot expect to determine
a sharp phase transition based on SCBA since SCBA does not describe
the localized phase of 1D metals. The disorder-induced topological
superconducting phase transition in spinless $p$-wave superconductors
occurs when the superconducting coherence length becomes comparable
to the localization length. In summary, SCBA is found to describe
qualitatively the suppression of the Majorana ZBP despite the fact
that it smears out the phase transition into a crossover from a topological
superconducting to a diffusive metallic phase. 
\begin{acknowledgments}
We acknowledge useful discussions with V. Stanev and V. Galitski.
This work is supported by JQI-NSF-PFC and Microsoft Q.
\end{acknowledgments}
\appendix

\section{SCBA in a Linearized Model\label{App:disorder-linear}}

The self-energy due to ensemble-averaged disorder is 
\begin{equation}
\Sigma\left(x,x'\right)=\delta\left(x-x'\right)\left\langle V\left(x\right)V\left(x'\right)G^{(0)}\left(x,x'\right)\right\rangle ,
\end{equation}
where $G^{(0)}$ is the Green function of the unlinearized fermion
operator. With the linearization $\psi\left(x\right)\simeq\psi_{R}e^{ik_{F}x}+\psi_{L}e^{-ik_{F}x}$,
$V$ and $G^{(0)}$ becomes
\begin{eqnarray}
G\left(x,x'\right) & \simeq & G_{RR}e^{ik_{F}\left(x-x'\right)}+G_{RL}e^{ik_{F}\left(x+x'\right)}\nonumber \\
 &  & +G_{LR}e^{-ik_{F}\left(x+x'\right)}+G_{LL}e^{-ik_{F}\left(x-x'\right)},\\
V\left(x\right) & \simeq & V_{f}\left(x\right)+V_{b}\left(x\right)e^{2ik_{F}x}\nonumber \\
 &  & +V_{b}^{*}\left(x\right)e^{-2ik_{F}x}.
\end{eqnarray}
Using the correlations given in Eqs.~(\ref{eqs:disorder-strengths}),
the self-energy becomes\begin{widetext}
\begin{eqnarray}
\Sigma\left(x,x'\right) & = & \delta\left(x-x'\right)\left\{ D_{f}G\left(x,x'\right)\right.\nonumber \\
 &  & +D_{b}\left[G_{RR}e^{ik_{F}\left(3x-3x'\right)}+G_{RR}e^{ik_{F}\left(-x+x'\right)}+G_{RL}e^{ik_{F}\left(3x-x'\right)}G_{RL}e^{ik_{F}\left(-x+3x'\right)}\right.\nonumber \\
 &  & \left.\left.+G_{LR}e^{-ik_{F}\left(3x-x'\right)}+G_{LR}e^{-ik_{F}\left(-x+3x'\right)}+G_{LL}e^{-ik_{F}\left(3x-3x'\right)}+G_{LL}e^{-ik_{F}\left(-x+x'\right)}\right]\right\} ,\\
 & \simeq & \delta\left(x-x'\right)\left\{ D_{f}G\left(x,x'\right)+D_{b}\left[G_{RR}e^{-ik_{F}\left(x-x'\right)}+G_{LL}e^{ik_{F}\left(x-x'\right)}\right]\right\} ,
\end{eqnarray}
\end{widetext}where in the last step only terms proportional to $e^{\pm ik_{F}x}$
are retained. The linearized same-point self-energy is therefore\begin{subequations}
\begin{eqnarray}
\Sigma_{RR} & = & D_{f}G_{RR}+D_{b}G_{LL},\\
\Sigma_{RL} & = & D_{f}G_{RL},\\
\Sigma_{LR} & = & D_{f}G_{LR},\\
\Sigma_{LL} & = & D_{f}G_{LL}+D_{b}G_{RR}.
\end{eqnarray}
\end{subequations}

When expressed in the chiral Nambu-Gorkov basis $\left(\psi_{R},\psi_{L},\psi_{L}^{\dagger},\psi_{R}^{\dagger}\right)$
used in the main text, we have
\begin{equation}
\Sigma=D_{f}\tau_{3}G\tau_{3}+\frac{D_{b}}{2}\tau_{3}\left(\sigma_{1}G\sigma_{1}+\sigma_{2}G\sigma_{2}\right)\tau_{3}.
\end{equation}

\section{Singularity of DOS at $\left(x=0,\omega=0\right)$ for $\Delta\ll\tau^{-1}$\label{App:pole}}

In the limit $\Delta\ll\tau^{-1}$, we treat $\Delta$ as a small
perimeter and expand the solution to Eq.~(\ref{eqs:ODE}) perturbatively
in $\Delta$. For simplicity we shall consider only Eq.~(\ref{eqs:ODE}a-c)
supplemented with the boundary conditions Eqs.~(\ref{eqs:BCbulk})
and Eq.~(\ref{eq:BCend-g02}), since the other equations are decoupled
and does not affect $g_{33}$ which determines the DOS. At $\Delta=0$
the problem is trivially solved with\begin{subequations}
\begin{eqnarray}
g_{33}^{(0)} & = & 1,\\
g_{31}^{(0)}=g_{02}^{(0)} & = & 0.
\end{eqnarray}
\end{subequations}

With small $\Delta$, we write $g_{J}=\sum_{n=0}^{\infty}g_{J}^{(n)}\Delta^{n}$
(for $J=\left\{ 33,31,02\right\} $) and expand Eq.~(\ref{eqs:ODE})
to successive orders in $\Delta$. To the first order in $\Delta$,
the system of differential equations is\begin{subequations}
\begin{eqnarray}
v_{F}\partial_{x}g_{31}^{(1)} & = & 2\omega g_{02}^{(1)}+2i+\frac{2i}{\tau}g_{02}^{(1)},\\
v_{F}\partial_{x}g_{02}^{(1)} & = & -2\omega g_{31}^{(1)},\\
v_{F}\partial_{x}g_{33}^{(1)} & = & 0,
\end{eqnarray}
\end{subequations}subjected to the boundary conditions of $g_{02}^{(1)}\left(0\right)=0$
and $\lim_{x\rightarrow\infty}g_{31}^{(1)}\left(x\right)=0$. This
is solved with \begin{subequations}
\begin{eqnarray}
g_{31}^{(1)}\left(x\right) & = & -\frac{e^{-2xi\sqrt{\omega\left(\omega+i\tau^{-1}\right)}/v_{F}}}{\sqrt{\omega\left(\omega+i\tau^{-1}\right)}},\\
g_{02}^{(1)}\left(x\right) & = & \frac{ie^{-2xi\sqrt{\omega\left(\omega+i\tau^{-1}\right)}/v_{F}}}{\omega+i\tau^{-1}}-\frac{i}{\omega+i\tau^{-1}},\\
g_{33}^{(1)}\left(x\right) & = & 0,
\end{eqnarray}
\end{subequations}which has no effect on the DOS. We must therefore
go to the second order which gives\begin{subequations}
\begin{equation}
v_{F}\partial_{x}g_{33}^{(2)}=-2ig_{31}^{(1)}-\frac{2i}{\tau}g_{31}^{(1)}g_{02}^{(1)},
\end{equation}
\end{subequations}where only the equation for $g_{33}^{(2)}$ is
given as it is relevant to the evaluation to DOS. Requiring $\lim_{x\rightarrow\infty}g_{33}^{(2)}\left(x\right)=\frac{1}{2\left(\omega+i\tau^{-1}\right)^{2}}$
which follows from the expansion of Eq.~(\ref{eq:g33bulk}), we have
\begin{eqnarray}
g_{33}^{(2)}\left(x\right) & = & -\frac{ie^{-4ix\sqrt{\omega\left(\omega+i\tau^{-1}\right)}/v_{F}}}{2\omega\tau\left(\omega+i\tau^{-1}\right)^{2}}-\frac{e^{-2ix\sqrt{\omega\left(\omega+i\tau^{-1}\right)}/v_{F}}}{\left(\omega+i\tau^{-1}\right)}\nonumber \\
 &  & +\frac{1}{2\left(\omega+i\tau^{-1}\right)^{2}},\\
g_{33}^{(2)}\left(0\right) & \approx & \frac{i\tau}{2\omega}-\frac{\tau^{2}}{2}-\frac{i\omega\tau^{3}}{2},
\end{eqnarray}
in which an expansion in $\omega$ is performed. We therefore see
that the pole at zero energy is present even for $\Delta\tau\ll1$.

\vfill{}

\bibliographystyle{apsrev4-1}
\bibliography{EilenbergerP}

\end{document}